\documentclass{article}
\usepackage[utf8]{inputenc}
\usepackage[english]{babel}
\usepackage{graphicx}
\usepackage{subcaption}
\usepackage{amssymb}
\usepackage{dirtytalk}
\usepackage{listings}
\usepackage{csquotes}
\usepackage{amsmath}
\usepackage{amsthm}
\usepackage{algpseudocode} 
\usepackage{algorithm} 
\usepackage{xcolor}
\usepackage{dirtytalk}
\usepackage{authblk}

\usepackage[margin=0.9in]{geometry}

\usepackage{comment} 

\usepackage[
backend=biber,
style=numeric,
maxnames=19,
sorting=none,
citestyle=numeric
]{biblatex}
\title{Economical-Epidemiological Analysis of the Coffee Trees Rust Pandemic}

\author[1,2,*]{Teddy Lazebnik}
\author[3]{Ariel Rosenfeld}
\author[4]{Labib Shami}

 \affil[1]{Department of Mathematics, Ariel University, Ariel, Israel}
 \affil[2]{Department of Cancer Biology, Cancer Institute, University College London, London, UK}
 \affil[3]{Department of Information Science, Bar-Ilan University, Ramat-Gan, Israel}
\affil[4]{Department of Economics, Western Galilee College, Acre, Israel}
\affil[*]{Corresponding author: lazebnik.teddy@gmail.com}

\date{}

\begin{document}

\maketitle

\begin{abstract}

Coffee leaf rust is a prevalent botanical disease that causes a worldwide reduction in coffee supply and its quality, leading to immense economic losses. While several pandemic intervention policies (PIPs) for tackling this rust pandemic are commercially available, they seem to provide only partial epidemiological relief for farmers. In this work, we develop a high-resolution economical-epidemiological model that captures the rust pandemic's spread in coffee tree farms and its associated economic impact. Through extensive simulations for the case of Colombia, a country that consists mostly of small-size coffee farms and is the second-largest coffee producer in the world, our results show that it is economically impractical to sustain any profit without directly tackling the rust pandemic. Furthermore, even in the hypothetical case where farmers perfectly know their farm's epidemiological state and the weather in advance, any rust pandemic-related efforts can only amount to a limited profit of roughly \(4\%\) on investment. In the more realistic case, any rust pandemic-related efforts are expected to result in economic losses, indicating that major disturbances in the coffee market are anticipated. \\ \\ 

\noindent

\noindent
\textbf{keywords:} coffee leaf rust, economical-epidemiological analysis, botanical pandemic control, pandemic intervention policies. 
\end{abstract}

\section{Introduction}
\label{sec:introduction}
Coffee is one of the most widely traded agricultural commodities in the world, second only to oil. The global coffee trade is estimated to support the livelihoods of around 100 million people around the world \cite{first_1}. In 2014 alone, an estimated 26 million farmers from 52 countries cultivated more than 8.5 million tons of coffee, accruing a value of almost \$39 billion in those countries \cite{first_2}. The retail value of coffee is significantly higher, in the United State during 2019 alone, sales reached \$87 billion \cite{first_3}. Smallholder farmers, typically with landholdings of 5 hectares or less, dominate production across most of the main cultivation regions \cite{intro_smallholders_1,intro_smallholders_3}. Specifically, In Colombia, the second-largest coffee producer in the world and the focus of this study, smallholder farms are the main coffee growing method \cite{eco_5}.

In recent years, coffee production has endured great disturbances due to climate change \cite{climate_cite}, economic shifts \cite{intro_smallholders_2}, and mostly importantly - the Coffee Leaf Rust (CLR) pandemic \cite{intro_2}. CLR is a destructive fungal disease that affects coffee plants, caused by the fungus Hemileia vastatrix and is one of the most significant threats to the global coffee industry, as it can significantly reduce crop yields and the quality of coffee beans \cite{intro_1}. It is one of the most widespread diseases of Coffea arabica in the world and the only significant coffee disease with a global distribution \cite{first_4,first_5}. In Colombia alone, coffee production decreased by 31\% between the years 2008 to 2011 due to the CLR outbreak compared to 2007. Similarly, in Central America as a whole, a 16\% decrease in coffee production was recorded for the 2012–2013 harvest, compared with 2011–2012 due to the rust pandemic. 

Currently, there is no \textit{silver bullet} solution for tackling the CLR pandemic despite several pandemic intervention policies (PIPs) which have shown promising results both in theory and in practice \cite{sec_3_4_chemical,sec_3_4_branch,sec_3_4_shade}. For example, some farmers have been using chemicals to fight the pathogen \cite{clr_pip_mexico} while others use various shading methods to stop the spread of the pathogen \cite{shading_2_3}. Unfortunately, despite the availability of these and similar PIPs, small coffee farms are still struggling to financially cope with the CLR pandemic. 

The main objective of this work is to determine whether an effective CLR management policy could bring about sustainable profit for small-size coffee farms. To that end, we develop a high-resolution model that captures the economical and epidemiological dynamics associated with the CLR pandemic. We instantiate our model for the case of Colombia and analyze the intelligent implementation of various popular PIPs through extensive simulations. 

The remainder of this paper is organized as follows: Section \ref{sec:related_work} provides a brief review of the epidemiological, economic, and PIP-related literature pertaining to CLR. Then, Section \ref{sec:model} formally introduces the proposed model, divided into the epidemiological, ecological, and economic dynamics and the interactions between them. Afterward, Section \ref{section:computer_implementation} introduces the simulation implementation of the proposed model using an agent-based simulation approach and a deep reinforcement learning-based agent to obtain near-optimal epidemiological-economic CLR management policies. Next, Section \ref{sec:experiments} presents our analysis using realistic configurations and historical data. Afterward, Section \ref{sec:discussion} discusses the results, draws conclusions, and suggests possible future work directions.

\section{Related work}
\label{sec:related_work}

Coffee is one of the world's most widely consumed beverages, with a global market that generates billions of dollars annually. Coffee production is a critical component of many economies, particularly in less-developed countries (LDCs), where it is a primary source of income for millions of small-scale farmers \cite{rw_intro_1,rw_intro_2,rw_intro_3}. This industry is critically influenced by the CLR pandemic, causing farms to financially collapse without external help \cite{rw_intro_4}. In recent years, several models have been developed to better understand the spread of the CLR pandemic and to inform the design and implementation of effective management strategies, also known as PIPs \cite{ref_1,snail_intro_solution,ref_4}. By using these models, researchers and policymakers can make informed decisions about the actions for controlling the CLR pandemic at either local, national, or international levels. Next, we briefly review the three main lines of work pertaining to our challenge: Botanical-Epidemiological models of CLR spread, PIPs for mitigating the CLR pandemic, and the economy of coffee production and CLR management.

\subsection{Botanical-Epidemiological Models}
Botanical-epidemiological models (BEMs) are special mathematical models that are used for the study of epidemiological dynamics in plant populations \cite{mathModelsPlantInfection, mathModelsPlantInfection2, EpidemicsPlantDiseasesMathematicalAnalysisAndModeling,review_svetlana}. These models are used to investigate factors that potentially contribute to the transmission and spread of diseases, as well as to predict the potential impacts of these diseases on crop yields and quality, food security, and many more \cite{human_pandemic_2, human_pandemic_3, human_pandemic_4, dynamical_analysis_of_fractional_plant_disease_model}. BEMs share common features and properties with animal and human-focused epidemiological models such as epidemiological states and infection mechanisms \cite{human_pandemic_1,nature_pandemic,cunniffe2015optimising}. As such, it is common to find BEMs that are based on the popular Susceptible-Infected-Recovered (SIR) model proposed by \cite{first_sir} for a human population. However, when tested on historical data, these SIR-based models demonstrate limited capabilities due to their over-simplicity and lack of consideration for the unique properties of the plant population and plant-based pathogens \cite{anggriani2017effect}.

Over time, researchers have been developing more complex and sophisticated BEMs that incorporate novel factors such as plant spatial distribution, host resistance, and pathogen virulence. 
For instance, \cite{second_sir} used Gaussian interaction and discretized SIR models to analyze disease spread in spatial populations with constant populations in 2-dimentional patches. Daily neighborhood interactions and contagion rates impact disease spread, with results indicating multiple waves with increasing size and a contagion rate determined by distance from the origin. Similarly, \cite{teddy_plants} used a spatio-temporal extended SIR epidemiological model with a non-linear output economic model to model the profit from a farm of plants during a botanical pandemic. In \cite{martcheva2009non}, the authors analyzed the stability, existence of a periodic solution, and coexistence of multiple strains in a multi-strain Susceptible-Infected-Susceptible (SIS) epidemic model. 
Generally speaking, extended SIR-based models, with unique properties to the pathogen, plant, and environment are taken into consideration for multiple scenarios, providing decent prediction capabilities \cite{motisi2022dark,FERRIS2019182,xf_host_resistance,plant_spatial_dist,host_resistance1,bebber2014global,pest_control_using_farming_awareness,sanya2022review,yu2022first,plants12030609,tschanz2022soybean,asian_soybean_rust_epidemics,cristancho2012outbreak, potato_late_blight_epidemiology}. 

The adoption of a model from one pathogen or plant to another is challenging due to the unique properties each combination of plant and pathogen has. Focusing on CLR dynamics, at a high level, the disease's local spread follows two stages that are similar to other diseases directly transmitted \cite{sec_2_3_1}. During the first stage, wind-carried urediniospores land on coffee farms and penetrate the stomata on the underside of the coffee leaves. Then, the urediniospores grow haustoria to extract nutrients from the leaf tissue, exiting again through the stomata and producing more spores. Pending, in the second stage, the urediniospores are dispersed to nearby coffee plants either through direct contact, water splash, or turbulent wind. It is also possible for the spores to be lifted into the atmosphere and contribute to the disease's spread in a larger area. 

The previous mathematical modeling work on CLR was limited compared to other types of plants and pathogens presumably due to the complexity associated with the CLR dynamics. Previous works include  \cite{empirical_multi_classifier_rust} where a machine learning-based model was proposed to classify CLR breakout from historical records. In \cite{in_silico_coffee_brt_model} the authors conducted in-silico experiments with the shading PIP, which is commonly implemented in banana trees, to predict the disease spread using a stochastic, nonlinear regression model (specifically, boosted regression trees model) to address questions regarding complex biological systems involving many variables and characterized by multiple interactions between processes. Back-propagation neural networks, nonlinear regression trees, and support vector regression models were used in \cite{two_level_classifier_rust} in a similar fashion.
However, common to these and similar works is the lack of attention to any economic aspects associated with the CLR and its mitigation. 

Aligned with prior work, in this study, we adopt an extended SIR-based BEM approach for the CLR pathogen. However, to the best of our knowledge, our proposed model is the most comprehensive to date by taking into consideration unique environmental factors such as rain, temperature, and humidity over time, in addition to the economic aspects which, as mentioned before, has yet to be considered in this context. 

\subsection{PIPs for CLR}

CLR-focused PIPs often seek to tackle the infection paths that the CLR pandemic's pathogen exploits \cite{sec_3_2_2}. In particular, three main PIPs are shown to be effective across a large number of farms and over time: 1) spraying with chemical substances; 2) shading; and 3) branch cutting. These PIPs, separately and combined, are the focus of this study. 

Chemicals are spread across coffee trees and react with the pathogen, eliminating it in the process. According to the Agriculture Ministry of Brazil, there are over 100 chemicals available to control CLR with different levels of aggressiveness and ecological side effects\footnote{We refer the interested reader to \url{http://www.agricultura.gov.br/servicos-e-sistemas/sistemas/agrofit}}. Indeed, \cite{sec_3_2_3} describes a four-year experiment of spraying chemical substances on a single farm in Brazil. The authors show that the effectiveness of this PIP is strongly dependent on the weather in general and the amount of rain in particular. However, on average, chemicals were able to stop a CLR outbreak in an efficient manner. Similar results were obtained in Mexico where \cite{clr_pip_mexico} showed that farmers who used chemicals to deal with a CLR outbreak were, on average, able to stop it. 

Shading is a completely different approach where one seeks to intercept rainfall which is one of the main infection paths used by the rust's pathogen. Namely, if rainfall is reduced around the coffee trees, the raindrops do not contact the trees directly and therefore do not contribute to the spore release of the pathogen. There are two main ways to implement the shading PIP \cite{belachew2020altitude, clr_pip_mexico, lopez2012shade, clr_shading_effect, shaded_and_unshaded_coffee_review}: synthetic and natural. For the synthetic solution, the shading is conducted by building shelters, usually using a dense net over the coffee trees \cite{shading_2_3}. The natural solution involves seeding higher trees such as banana trees that naturally shade the coffee trees \cite{in_silico_coffee_brt_model}. While the former solution is more expensive and slows the growth of young trees due to the lack of sun \cite{light_intenties_effect}, the latter has other shortcomings as well. If rainfall is heavy, the shade trees form gutters, where water builds up into larger drops. These drops operate as a more efficient infection vector for the CLR \cite{tree_shading, intro_solutions, cropping_effects_on_rust, shaded_and_unshaded_coffee_review}. 

Lastly, implementing the branch cutting PIP can reduce the infection rate of CLR \cite{cropping_effects_on_rust}. Nonetheless, the influence of coffee tree branch pruning is unclear as certain cropping practices affect the host or microclimate in different ways, which can either favor or hamper different aspects of the infection cycle. For instance, \cite{shade_management_and_pruning} analyzed the impact of shade tree management and pruning on CLR in two coffee varieties in the Peruvian Amazon. The authors found that coffee plants exposed to polyculture on the side and with their stems cut at 40 cm from the ground had lower incidences and severity of CLR. Another study by \cite{pruning_in_Hawai} recommended using pruning methods along with fungicides to decrease CLR infection in non-resistant coffee varieties, showing that pruning alone obtains poor results.

It is important to note that a farmer may implement any of the above PIPs, separately or combined, for any desired time frame and to different extents. Namely, for a given production period, a farmer need not necessarily \say{commit} to a single PIP setup. 

\subsection{The economy of coffee production}

Like all business owners, coffee farm owners seek to maximize their profits. Since smallholder farms dominate the coffee production market, each farm owner is assumed to have a negligible effect on the world price \cite{mas1998theory}. As such, farmers are assumed to consider the unit price to be given. Accordingly, the main determinant of the competitiveness of individual farms in the world market is the cost of production at the farm level.

Following \cite{intro_intro_1}, the coffee production costs can be largely disentangled into four primary categories: paid labour (i.e., formal workers), unpaid labour (i.e., informal workers such as family members), inputs (i.e., herbicides, pesticides, and fertilizer), and fixed costs (i.e., installation costs, finance costs, depreciation, and such). 
CLR controlling measures are implemented by most, or all, coffee farmers, hence they are considered part of the farm's fixed costs. Conversely, CLR-elimination costs are added to the event of a CLR outbreak. 

Let us consider the recent cost structure and break-even points analysis in different coffee origins as examined by Fairtrade USA and Cornell University\footnote{\url{https://www.roastmagazine.com/resources/Articles/2017/2017_Issue3_MayJune/Roast_MayJune17_FinSustCoffeeProd.pdf}}. The authors have estimated the costs of smallholder coffee farms in Honduras, Peru, Colombia, and Mexico. Using average costs and productivity, they construct a \say{representative} producer for each cooperative. They then utilized this  \say{representative} farmers to calculate four break-even points: one that considers only variable costs; one that adds fixed costs; one that includes depreciation; and one that accounts for the amortization of farm establishment costs, land costs, labor costs, and physical capital. They concluded farm owners in all studied origins face great uncertainty as to their long-term viability.
To the best of our knowledge, the most comprehensive study on the matter at hand is \cite{intro_intro_1}. The authors use an original large-scale dataset comprising cross-sectional data from three Arabica-producing countries in Latin America: Colombia, Costa Rica, and Honduras. The study examines farm level data that allows an investigation of the distribution of costs and profitability across farms in these three important coffee origins. 
The analysis demonstrated the high heterogeneity and variability across individual farm owners.

From an economical perspective, CLR is considered to be the most destructive disease affecting coffee in the world, affecting both yield quality and quantity \cite{intro_intro_4}. Thus, for farmers, it is economically disastrous and can result in up to 50 percent yield loss \cite{sera2022coffee}.
With climate change, the CLR has become even more damaging, even in areas that were previously known to be less prone to the disease \cite{intro_intro_2}. According to the Tropical Agricultural Research and Higher Education Center, the impact of the CLR attack on the Central America region in 2012-2013 caused a 15\% reduction in crop production on average \footnote{\url{https://www.researchgate.net/profile/Elias-De-Melo-Virginio-Filho/publication/340494371_Prevention_and_control_of_coffee_leaf_rust_Handbook_of_best_practices_for_extension_agents_and_facilitators_Prevention_and_control_of_coffee_leaf_rust_Handbook_of_best_practices_for_extension_agents_a/links/5e8d22bca6fdcca789fde44c/Prevention-and-control-of-coffee-leaf-rust-Handbook-of-best-practices-for-extension-agents-and-facilitators-Prevention-and-control-of-coffee-leaf-rust-Handbook-of-best-practices-for-extension-agents-a.pdf}}. Honduras and El Salvador were the countries that experienced the highest percentage of crop loss - 31\% and 23\%, respectively.
It is also important to note that CLR outbreaks, similar to other production disturbances such as the 2008 global financial crisis and the COVID-19’s socioeconomic disruptions, have been linked to a significant reduction in investment in coffee farms \cite{intro_intro_3} and increased unemployment \cite{cristancho2012outbreak}.

\section{Model}
\label{sec:model}

The proposed model consists of three interconnected components: a BEM component that describes the CLR pandemic spread at both tree and farm levels while considering the ecological dynamics that describe the temporal environmental changes; an economic component that defines the farmers' expenses and revenues; and a PIP component that specifies which, if any, PIPs the farmer implements. Each of these components, as well as the interactions between them, are detailed below.

\subsection{BEM component}
In order to capture the spread of the CLR pandemic, we utilize a multi-scale BEM which models the pandemic's spread at both the individual tree and farm levels as well as the temporal environmental dynamics. At the tree level, we use an ordinary differential equation (ODE) approach, indicating the epidemiological state and disease dynamics between each of the tree's branches. 
This granularity level was shown to well explain the CLR pandemic spread in individual coffee trees in the past (see \cite{ref_4}). As such, a more fine-grained resolution, such as considering each leaf separately, would likely not contribute much to capturing the actual dynamics while introducing a seemingly unnecessary, yet non-negligible, computational burden. At the farm level, we define the CLR spread between trees of the same farm using a spatio-temporal approach, taking into consideration the distances between the trees. Distances between trees are well known to play a central role in disease spread \cite{mathModelsPlantInfection2}. Last, an ecological part, representing the temporal environmental dynamics that are unrelated to the CLR dynamics, is applied as well. The three parts of the BEM are discussed next.  

\subsubsection{Tree level spread}
\begin{figure}[hbt!]
    \centering
    \includegraphics[width=0.99\textwidth]{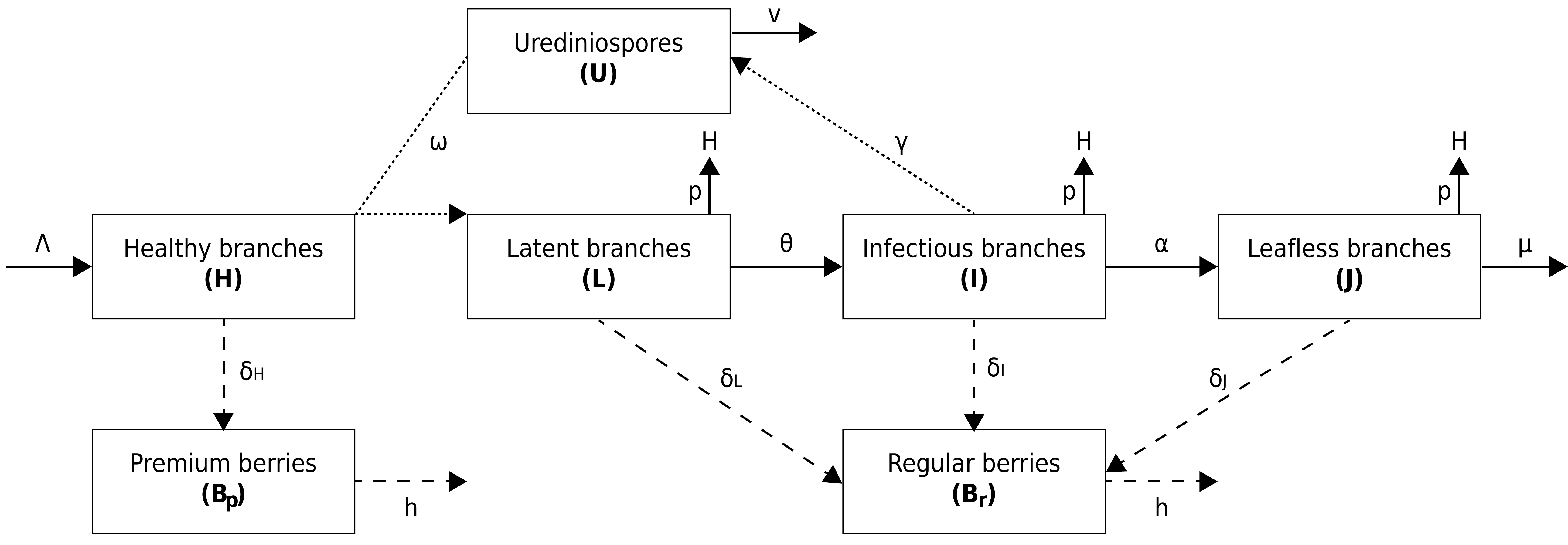}
    \caption{A schematic illustration of the tree level CLR spread dynamics. Solid lines indicate changes in the epidemiological state of the coffee tree's branches. Dashed lines indicate berries production. Dotted lines indicate the production and interaction of urediniospores with the coffee tree's branches.}
    \label{fig:tree_scheme}
\end{figure}

The coffee tree is a perennial plant that has persistent leaves. The annual productivity period lasts around nine months and strongly depends on the coffee species' cultivars and climate conditions. Here, we describe the evolution of the number of coffee branches according to their epidemiological state corresponding to the CLR life cycle: healthy, latent (i.e., pathogen-exposed but not contagious), infectious, and leafless. The infection of a branch corresponds to the infection of all its leaves. Formally, the pandemic spread on the tree level takes the following system of ODEs' form:

\begin{equation}
    \begin{array}{l}
        \frac{d H(t)}{d t} = \Lambda(t) - \omega(t) U(t)H(t) + p_L(t)L(t) + p_I(t)I(t) + p_J(t)J(t), \\ \\
        
        \frac{d L(t)}{d t} = \omega(t) U(t)H(t)  - \theta (t) L(t) - p_L(t)L(t), \\ \\
        
        \frac{d I(t)}{d t} = \theta (t) L(t) - \alpha(t) I(t) - p_I(t)I(t), \\ \\
        
        \frac{d J(t)}{d t} = \alpha(t) I(t) - \mu(t) J(t) - p_J(t)J(t), \\ \\
        
        \frac{d U(t)}{d t} = \gamma (t) I(t) - v(t) U(t) - p_U(t) U(t), \\ \\
        
        \frac{d B_p(t)}{d t} = \delta_H(t) H(t) - h(t), \\ \\
        
        \frac{d B_r(t)}{d t} = \delta_L(t) L(t) + \delta_I(t) I(t) + \delta_J(t) J(t) - h(t), 
    \end{array}
    \label{eq:tree_level_system}
\end{equation}
where \(H(t), L(t), I(t), J(t), U(t), B_p(t)\) and \(B_r(t)\) represent the number of healthy, latent, infectious and leafless branches, urediniospores, premium berries, and regular berries, respectively, at time \(t\). The total quantity of branches of the tree is denoted by \(N(t) := S(t) + L(t) + I(t) + J(t)\). A schematic illustration of the tree level pandemic spread model is shown in Fig. \ref{fig:tree_scheme}. A full description of Eq. (\ref{eq:tree_level_system}) is provided below.

The coffee tree is a persistent-leaved perennial plant whose annual productivity is influenced by factors such as species, cultivars, and climate conditions. The proposed model is based on the coffee leaf rust life cycle, including states of healthy, latent, infectious, and leafless branches. The evolution of the pandemic is represented by a system of ODEs, taking into account the number of branches in each epidemiological state and their interactions over time. The model represents the amount of healthy (\(H\)), latent (\(L\)), infectious \((I\)), and leafless branches (\(J\)), as well as urediniospores (\(U\)), premium berries \((B_p)\), and regular berries \((B_r)\) at a given time \((t)\), along with the total number of branches in the tree. The epidemiological dynamics are described in Eq. (\ref{eq:12}-\ref{eq:18}).

In Eq. (\ref{eq:12}) \(\frac{d H(t)}{d t} \) is the dynamic number of healthy branches. It is affected by the following five terms. First, the recruitment of healthy branches at rate \(\Lambda(t)\). Second, the number of healthy branches that become latent per spore deposited, known as germination efficacy, at rate \(\omega\). Third, the impact of the PIP on latent branches, \(p_L\), makes them healthy again. Fourth, the impact of the pip on infectious branches, \(p_I\), makes them healthy again. Lastly, the impact of the pip on leafless branches, \(p_J\), makes them healthy again. 

\begin{equation}
    \frac{d H(t)}{d t} =  \Lambda(t) - \omega(t) U(t)H(t) + p_L(t)L(t) + p_I(t)I(t) + p_J(t)J(t).
    \label{eq:12}
\end{equation}

In Eq. (\ref{eq:13}), \(\frac{d L(t)}{d t} \) is the dynamic number of the latent branches of a coffee tree. It is affected by the following three terms. First, healthy branches covered in spores change into latent branches at a rate \(\omega(t)\). Second, latent branches become infectious at rate \(\theta\). Lastly, the impact of the PIP on latent branches, \(p_L\), makes them healthy again. 

\begin{equation}
    \frac{d L(t)}{d t} =  \omega(t) U(t)H(t)  - \theta (t) L(t) - p_L(t)L(t).
    \label{eq:13}
\end{equation}

In Eq. (\ref{eq:14}), \(\frac{d I(t)}{d t}\) is the dynamic number of infected branches of a coffee tree. It is affected by the following three terms. First, the latent branches become infectious at rate \(\theta(t)\) where \(\frac{1}{\theta(t)}\) corresponds to the latency period. Secondly, infected branches become leafless at a rate of \(\alpha(t)\), where \(\frac{1}{\alpha(t)}\) is the sporulation period. Lastly, the impact of the PIP on infected branches \(p_I\), makes them healthy again.

\begin{equation}
    \frac{d I(t)}{d t} = \theta (t) L(t) - \alpha(t) I(t) - p_I(t)I(t).
    \label{eq:14}
\end{equation}

In Eq. (\ref{eq:15}), \(\frac{d J(t)}{d t}\) is the dynamic number of leafless branches of a coffee tree. It is affected by the following three terms. First, infected branches become leafless at a rate of \(\alpha(t)\), where \(\frac{1}{\alpha(t)}\) is the sporulation period.
The second factor is natural mortality, which has a baseline rate \(\mu(t)\) for all branches
Lastly, the impact of the PIP on leafless branches \(p_I\), that makes them healthy again

\begin{equation}
    \frac{d J(t)}{d t} = \alpha(t) I(t) - \mu(t) J(t) - p_J(t)J(t).
    \label{eq:15}
\end{equation}

In Eq. (\ref{eq:16}), \(\frac{d U(t)}{d t}\) is the dynamic number of urediniospores branches of a coffee tree. It is affected by the following three terms. First, a urediniospore branch is produced when an infection branch is produced at a rate of \(\gamma\). Second, urediniospore branches are deposited at a rate of \(v(t)\). Additionally, the PIP's impact on urediniospore branches \(p_I\) makes them healthy again.

\begin{equation}
    \frac{d U(t)}{d t} = \gamma (t) I(t) - v(t) U(t) - p_U(t) U(t).
    \label{eq:16}
\end{equation}

The quantity of harvested berries, \(h\) is also important as it is considered in the equations and it affects the total amount of berries that can be harvested, although it does not have a direct impact on the dynamics of premium berries. In (\ref{eq:17}) \(\frac{d B_p(t)}{d t}\) is the amount of premium berries. It is impacted by the following terms. A healthy branch creates premium berries at rate \(\delta_H\). Secondly, berries are harvested at a rate of h.

\begin{equation}
    \frac{d B_p(t)}{d t} = \delta_H(t) H(t) - h(t).
    \label{eq:17}
\end{equation}

In Eq. (\ref{eq:18}), \(\frac{B_r(t)}{dt}\) represents the change in the number of regular berries over time. This quantity is influenced by four factors: the rate at which berries are produced when branches are dormant \(\delta_L\), the rate at which berries are produced when branches are infected \(\delta_I\), the number of berries that were produced while there were leafless branches \(\delta_J\), and \(h\) the number of berries that have already been harvested.

\begin{equation}
    \frac{d B_r(t)}{d t} = \delta_L(t) L(t) + \delta_I(t) I(t) + \delta_J(t) J(t) - h(t).
    \label{eq:18}
\end{equation}

The tree level epidemiological model's parameters and their descriptions are summarized in Table \ref{table:tree_level_params}. In addition, in the same table, the average parameter value and the corresponding source are provided.

\begin{table}[!ht]
\centering
\begin{tabular}{p{0.1\textwidth}p{0.4\textwidth}p{0.15\textwidth}p{0.15\textwidth}}
\hline
Parameter & Biological meaning & Average value & source \\ \hline
 \(\Lambda\) & Recruitment rate \([day^{-1}]\) & 7 & \cite{ref_4} \\
 \(\omega\) & Inoculum effectiveness \([1]\) & 5 & \cite{rayner1961germination}\\
 \(p_L\) & PIP influence rate on lantent branches \([1]\)  & NA & NA \\
 \(p_I\)& PIP influence rate on infected branches \([1]\) & NA & NA \\
 \(p_J\) & PIP influence rate on leafless branches \([1]\) & NA & NA \\
 \(\theta\) & Infection rate  \([day^{-1}]\) &  25.5 & \cite{waller1982coffee} \\
 \(\alpha\) & Infected branches become leafless rate \([day^{-1}]\) & 150 & \cite{waller1982coffee} \\
 \(\mu\) & Natural mortality rate \([day^{-1}]\) & 0.0084 & \cite{ref_4} \\
 \(\gamma\) & Urediniospore production from infections branches rate \([day^{-1}]\) & \(10\) & \cite{urediniospore-productio-avg-rate} \\
 \(v\) & Deposition rate \([day^{-1}]\) &  0.09 & \cite{urediniospore-productio-avg-rate} \\
 \(p_U\) & Urediniospore deposition rate \([m^2 day^{-1}]\) & 5000 & \cite{ref_5} \\
 \(\delta_H\) & Berries' production rate by healthy branches \([day^{-1}]\) & 0.35 & \cite{ref_4} \\
\(\delta_L\)&  Berries' production rate by latent branches \([day^{-1}]\) & 0.25 &  \cite{ref_4}\\
\(\delta_I\) & Berries' production rate by infected branches \([day^{-1}]\) & 0.15 & \cite{ref_4} \\
\(\delta_J\) & Berries' production rate by leafless branches \([day^{-1}]\) & 0.025 & \cite{ref_4} \\
h & Harvesting rate \([1]\) & NA & NA \\
\end{tabular}
\caption{Description for the parameters used in the tree level rust-pandemic spread (see Eq. (\ref{eq:tree_level_system})). NA stands for historically or empirically non-available data that is picked for each instance separately.}
\label{table:tree_level_params}
\end{table}

\subsubsection{Farm level spread}
\begin{figure}[hbt!]
    \centering
    \includegraphics[width=0.85\textwidth]{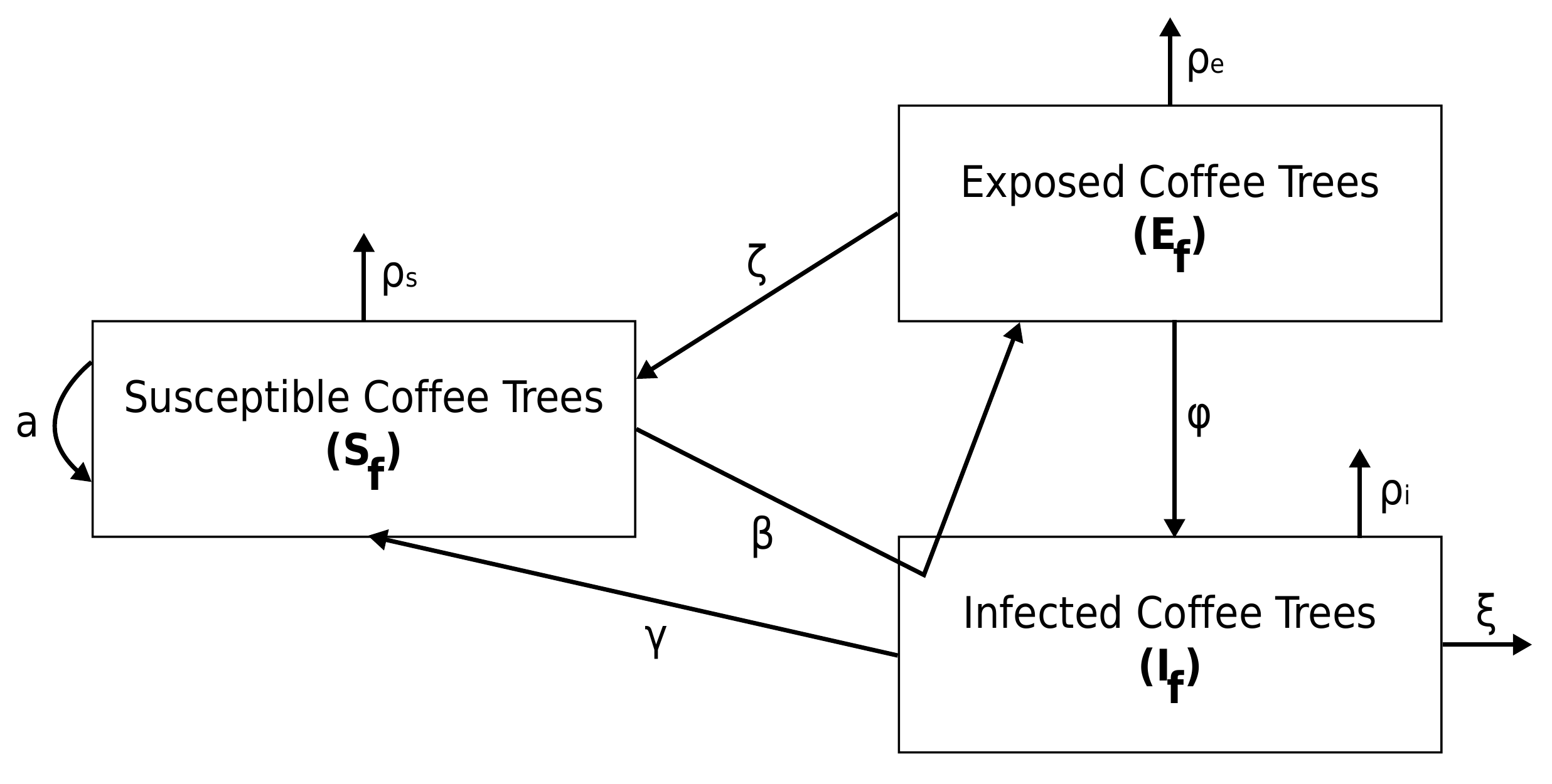}
    \caption{A schematic view of the farm level CLR pandemic dynamics. }
    \label{fig:farm_scheme}
\end{figure}

For the farm level spread dynamics, we extended the spatio-temporal SEIS  model \cite{seirs_1,seirs_2,seirs_3} with the following three modifications: 1) we assume susceptible coffee trees can be added to the farm over time (as observed in \cite{farm_start_1}); 2) coffee trees of all epidemiological states can die due to non-CLR-related reasons (see \cite{farm_start_2} for a discussion); and 3) exposed trees can become susceptible without becoming infection first (as noted by \cite{farm_start_3}). Formally, the pandemic spread at the farm level takes the following system of ODEs' form:
\begin{equation}
    \begin{array}{l}
         \frac{\partial S_f(t, x)}{\partial t} =  a(t) S_f(t, x) - \rho_s (t) S_f(t, x) - \beta(t, x) S_f(t, x) I(t, x) + \gamma(t) I_f(t, x) + \zeta(t) E_f(t), \\ \\
         
         \frac{\partial E_f(t, x)}{\partial t} =  \beta(t, x) S_f(t, x) I(t, x) - \phi(t) E_f(t, x) - \rho_e(t) E_f(t, x) - \zeta(t) E_f(t), \\ \\
         
         \frac{\partial I_f(t, x)}{\partial t} =  \phi(t) E_f(t, x) - \gamma(t) I_f(t, x) - \xi(t) I_f(t, x)  - \rho_i(t) I_f(t, x), \\ \\
         
    \end{array}
    \label{eq:farm_level_model}
\end{equation}
where \(S_f(t,x), E_f(t,x)\), and \(I_f(t,x)\)  represent the densities of susceptible, exposed, and infected coffee trees, respectively, at time \(t\) and location \(x\). A schematic illustration of the farm level spread dynamics is provided in Fig. \ref{fig:farm_scheme}. The farm level model construction is done as follows.

In Eq. (\ref{eq:farm_appendix_s}), \(\frac{dS_f(t, x)}{dt}\) is the dynamic density of susceptible CTs over time. It is affected by the following five terms. First, new susceptible CTs are born at rate \(a(t)\) with respect to the current amount of susceptible CTs. Second, susceptible CTs die out to non-pandemic related causes at a rate \(\rho_s (t)\). Third, susceptible CTs are infected by infectious CTs at a rate \(\beta (t)\) and become exposed. Forth, infectious CTs are recovered with a rate \(\gamma (t)\) and become susceptible CTs. Fifth, exposed CTs can become susceptible again if their non-healthy branches are removed before becoming infectious at a rate \(\zeta (t)\).  

\begin{equation}
         \frac{\partial S_f(t, x)}{\partial t} = a(t) S_f(t, x) - \rho_s (t) S_f(t, x) - \beta(t, x) S_f(t, x) I(t, x) + \gamma(t) I_f(t, x) + \zeta(t) E_f(t).
    \label{eq:farm_appendix_s}
\end{equation}

In Eq. (\ref{eq:farm_appendix_e}), \(\frac{dE_f(t, x)}{dt}\) is the dynamic density of exposed CTs over time. It is affected by the following four terms. First, susceptible CTs are infected by infectious CTs at a rate \(\beta (t)\) and become exposed. Second, exposed CTs become infectious at a rate \(\phi (t)\). Third, exposed CTs die out to non-pandemic related causes at a rate \(\rho_e (t)\). Forth, exposed CTs can become susceptible again if their non-healthy branches are removed before becoming infectious at a rate \(\zeta (t)\).  

\begin{equation}
          \frac{\partial E_f(t, x)}{\partial t} =  \beta(t, x) S_f(t, x) I(t, x) - \phi(t) E_f(t, x) - \rho_e(t) E_f(t, x) - \zeta(t) E_f(t).
    \label{eq:farm_appendix_e}
\end{equation}

In Eq. (\ref{eq:farm_appendix_i}), \(\frac{dI_f(t, x)}{dt}\) is the dynamic density of infectious CTs over time. It is affected by the following fourth terms. First, exposed CTs become infectious at a rate \(\phi (t)\). Second, infectious CTs either recover or die at a rate \(\gamma (t)\). Third, infected CTs die out of non-pandemic related causes at a rate \(\xi\).
Fourth, infectious CTs die out to non-pandemic related causes at a rate \(\rho_i (t)\).      
         
\begin{equation}
         \frac{\partial I_f(t, x)}{\partial t} =  \phi(t) E_f(t, x) - \gamma(t) I_f(t, x) - \xi(t) I_f(t, x) - \rho_i(t) I_f(t, x).
    \label{eq:farm_appendix_i}
\end{equation}

The farm level epidemiological model's parameters and their descriptions are summarized in Table \ref{table:farm_level_params}. 

\begin{table}[!ht]
\centering
\begin{tabular}{p{0.1\textwidth}p{0.6\textwidth}}
\hline
Parameter & Biological meaning  \\ \hline
 \(a\) &  Susceptible tree's rate of birth  \\
 \(\rho_s\) & Natural mortality rate of susceptible    \\
 \(\beta\) & The rate at which susceptible trees are exposed to the pathogen\\
 \(\gamma\) & The rate at which infected trees recover   \\
 \(\zeta \)& The rate at which exposed trees  recover due to some PIP \\
 \(\phi\) &  The rate at which exposed trees  become infectious \\
 \(\rho_e\) & Natural mortality rate of trees due to the pathogen \\
 \(\xi \)& The average death rate of infected trees due to the rust pandemic. \\
 \(\rho_i\) &  Natural mortality rate of infected trees due to the pathogen \\
 &   
\end{tabular}
\caption{Description for the parameters used in the farm level rust-pandemic spread (see Eq. (\ref{eq:farm_level_model})). }
\label{table:farm_level_params}
\end{table}

Note that the coefficients of Eq. (\ref{eq:farm_level_model}) strongly depend on several factors including the epidemiological states of the coffee trees on the farm, the temporal changes in the environment, and the PIPs implemented. First, the susceptible growth parameter, \(a(t)\), depends on the farmer's decision to seed more trees as we assume minimal to no natural reproduction taking place in an organized farm \cite{coffee_burn_1,coffee_burn_2,coffee_burn_3}. The average infection rate of the infectious coffee tree population (\(I_f\)) to the susceptible coffee tree population (\(S_f\)), \(\beta\), is proportional to the relative infection state of the infectious tree as defined by the portion of infectious and leafless branches the infectious coffee tree has and the opposite distances between the two coffee trees. Thus, \(\beta(t, x)\) is defined as follows:
\begin{equation}
    \beta(t, x) = \frac{\sum_{s \in S_f(t, x)} \sum_{i \in I_f(t, x)} \frac{\beta(t) i_{U}}{d(s, i)}}{S_f(t,x) * I_f(t, x)} ,  
    \label{eq:beta_farm}
\end{equation}
where \(d(s, i)\) is a metric function that gets two coffee trees on a farm and returns the distance between them (i.e., Euclidean distance), \(i_{x}\) is the \(x\) branch population of the \(i\)'s coffee tree, \(\beta(t)\) is the average infection rate of the pathogen at time \(t\), changing over time due to changes in the environmental conditions. The average recovery rate from the infection state and duration of the infectious coffee tree, \(\gamma(t)\), is proportional to the average amount of infectious coffee trees that their latent, infectious, and leafless branches are removed or recovered. Formally, \(\gamma(t)\) takes the form:
\begin{equation}
    \gamma(t) = \frac{\sum_{i \in I_f(t, x)} \sum_{x \in \{L, I, J\}} i_{x}(t) - i_{x}(t-1)}{I_f(t, x)}.  
    \label{eq:gamma_farm}
\end{equation}
The recovery process from the exposed state is similar, and therefore the recovery rate from the exposed state, \(\zeta(t)\), takes the form:
\begin{equation}
    \zeta(t) = \frac{\sum_{e \in E_f(t, x)} e_{L}(t) - e_{L}(t-1)}{E_f(t, x)}, 
    \label{eq:zeta_farm}
\end{equation}
where \(e_{L}\) is the number of latent branches of tree \(e\).
The coffee trees that do not recover during the exposed state become infected as part of their branches transform from latent to infection state. Thus, the average exposed to infection rate, \(\phi(t)\) is defined as follows:
\begin{equation}
    \phi(t) = \frac{\sum_{e \in E_f(t, x)} sign(e_{I}(t)*e_{I}(t-1))}{E_f(t, x)},
    \label{eq:phi_farm}
\end{equation}
where \(e_{I}\) is the number of infected branches of tree \(e\).
In a complementary manner, non-recovered infected coffee trees die out due to the CLR pandemic at an average rate, \(\xi(t)\), which depends on the amount of non-susceptible branches that coffee tree has and a threshold of the pathogen-carrying capability of the coffee tree before it dies:
\begin{equation}
    \xi(t) = \frac{\sum_{i \in I_f(t, x)} D(i_{L}, i_{I}, i_{J}) }{I_f(t, x)},
    \label{eq:xi_farm}
\end{equation}
where \(D\) is a function that accepts the coffee tree's number of latent, infected, and leafless branches and returns $1$ if the tree dies due to the disease and $0$ otherwise. 
Finally, \(p_{s}, p_{e}\), and \(p_{i}\) are the natural die-out rates of the susceptible, exposed, and infected coffee trees, respectively. These parameters are defined by natural causes not directly related to the CLR pandemic. 

\subsubsection{Ecological Dynamics}
The CLR spread, as well as the coffee trees' berries production, depend on multiple environmental and climate processes \cite{eco_1,eco_2,eco_3}. In particular, temperature, humidity, and rain are considered to be the primary ecological properties to have a direct influence on the CLR pandemic spread \cite{eco_4,eco_5,eco_6}. Formally, these three processes are represented using three time-series vectors \(C_T(t), C_H(t)\), and \(C_R(t)\), respectively. 

Due to the complexity of associating these three ecological processes with the tree level CLR pandemic spread, we utilize a machine learning approach to learn the association between the two. Formally, we define a function \(C: \mathbb{R}^{3} \rightarrow \mathbb{R}^{12}\) such that \(C(C_T, C_H, C_R) \rightarrow [\lambda, \omega, \theta, \alpha, \mu, \gamma, v, \delta_H, \delta_L, \delta_I, \delta_J, h]\) that maps the ecological processes to the tree level parameters. Specifically, in order to obtain \(C\), we used historical data from \cite{eco_data_1,eco_data_2,coffee_data} and the \textit{TPOT} automatic machine learning framework \cite{tpot} with the stratified k-fold cross-validation method \cite{k_fold_time_series}. 

\subsection{Economical component}
\label{sec:eco_model}
As discussed before, the costs of coffee farms, denoted by \(TC\), can be largely classified into four main categories (see Table \ref{table:cost_production}): paid labour (\(PL\)), unpaid labour (\(UL\)), inputs (chemical and organic) (\(IN\)), and fixed costs (\(FC\)); such that \(TC := FC + PL + UL + IN\). Notably, the first three components are depended on the number of trees on the farm. These are the basic costs, which do not take into consideration the additional cost associated with controlling a CLR outbreak. To this end, assuming that the CLR-elimination costs are only encountered if an outbreak occurs (i.e., \(I_f > 0\)), the additional cost of controlling the pandemic, \(SC\) is added the total cost, \(TC\), resulting in \(TC := FC + PL + UL + IN + SC\). Specifically, \(SC\) is defined by the PIPs utilized by the farmer, as defined in the next section.

\subsection{PIPs}
\label{sec:pips}
We consider three potential PIPs: 1) spraying chemical substances \cite{sec_3_4_chemical}; 2) shading  \cite{sec_3_4_shade}; and 3) branch cutting \cite{sec_3_4_branch}. 
Technically, each of these PIPs has a scale of usage defined by a specific parameterization space influencing both the epidemiological state of the farm and the economic state due to the cost associated with implementing each PIP. Next, we define these in detail. 

The spraying chemical substances (SCS) PIP is defined by the number of substances \(c \in \mathbb{R}\) applied per acre and the number of acres \(a \in \mathbb{R}\) the PIP is implemented in. The amount of substances reduces the amount of pathogen in each tree linearly at the time of implementation. Thus, Eq. (\ref{eq:tree_level_system}) takes the form: 
\[
\frac{d U(t)}{d t} = \gamma (t) I(t) - v(t) U(t) - p_U(t) U(t) - \kappa_{ca}c\frac{a}{A},
\]
where \(\kappa_{ca} \in \mathbb{R}^+\) is the efficiency of the active material and \(A \in \mathbb{R}^+\) is the total size of the farm. 
From an economic point of view, the SCS PIP has an associative cost that satisfies:
\begin{equation}
    scs_{cost}(t, [c, a]) = \left.
      \begin{cases}
         \big ( \kappa_1 log(a) + \kappa_2\big ) \cdot \big ( \kappa_3 log(c) + \kappa_4 \big ) & c \leq \kappa_c \wedge a \leq \kappa_a \\
         
       \big ( \kappa_1 log(a) + \kappa_2\big ) \cdot \big ( \kappa_3 log(\kappa_c) + \kappa^o_c (c - \kappa_c) + \kappa_4 \big ) & c > \kappa_c \wedge a \leq \kappa_a \\
       
        \big ( \kappa_1 log(\kappa_a) + \kappa^o_a (a - \kappa_a) + \kappa_2\big ) \cdot \big ( \kappa_3 log(c) + \kappa_4 \big ) & c \leq \kappa_c \wedge a > \kappa_a\\
        
        \big ( \kappa_1 log(\kappa_a) + \kappa^o_a (a - \kappa_a) + \kappa_2 \big ) \cdot( \kappa_3 log(\kappa_c) + \kappa^o_c (c - \kappa_c) + \kappa_4 \big )& c > \kappa_c \wedge a > \kappa_a
      \end{cases}
      \right\}    
    \label{eq:pip_scs}
\end{equation}
where \(\kappa_i \in \mathbb{R}\) such that \(i \in [1, \dots, 6] \wedge \{c, a\}\) are free parameters of the cost are defined by the external supply and demand over time. 
 
The shading PIP is designed to cover the coffee tree from raindrops that carry the pathogen between the branches. As such, the shading PIP influences the function \(C\) through the \(C_R\) function by reducing the average amount of \(C_R\) with respect to the coverage in acres of the farm, \(a \in \mathbb{R}^+\). Formally, for shading PIP with coverage \(a\) takes the form: 
\[
C_R \leftarrow C_R \frac{a}{A}.
\]
The cost associated with the shading PIP is the installation costs:
\begin{equation}
    shading_{cost}(t, [a]) = \left.
      \begin{cases}
         \kappa_1 log(a) + \kappa_2 & a \leq \kappa_a  \\
         
        \kappa_1 log(\kappa_a) + \kappa^o_a (a - \kappa_a) + \kappa_3 & a > \kappa_a.
      \end{cases}
      \right\}    
    \label{eq:pip_sct}
\end{equation}

Finally, the branch cutting (BC) PIP is simply cutting \(\zeta \in \mathbb{N}\) branches. Initially, leafless branches are cut at random, as these are easy to detect. Afterward, if \(\zeta\) is larger than the number of the leafless branches in the CT, the number of non-leafless branches is pruned at random.

Namely, 
\begin{equation}
    \begin{array}{l}
    \frac{d H(t)}{d t} = -max(0, \zeta - J(t))\frac{H(t)}{H(t) + L(t) + I(t)}, \\ \\
        
        \frac{d L(t)}{d t} = -max(0, \zeta - J(t))\frac{L(t)}{H(t) + L(t) + I(t)}, \\ \\
        
        \frac{d I(t)}{d t} = -max(0, \zeta - J(t))\frac{I(t)}{H(t) + L(t) + I(t)}, \\ \\
        
        \frac{d J(t)}{d t} = -min(\zeta, J(t)).
    \end{array}
        \label{eq:bcct_epi}
\end{equation}
The cost associated with the BC PIP is the manpower required to cut branches across the farm's grounds. As such, for \(a \in \mathbb{R}\) acres and \(\zeta\) cutting effort per tree, the costs take the form:
\begin{equation}
    BC_{cost}(t, [\zeta, a]) = \left.
      \begin{cases}
         \big ( \kappa_1 log(a) + \kappa_2\big ) \cdot \big ( \kappa_3 \zeta + \kappa_4 \big ) & a \leq \kappa_a \\

        \big ( \kappa_1 log(\kappa_a) + \kappa^o_a (a - \kappa_a) + \kappa_2\big ) \cdot \big ( \kappa_3 \zeta + \kappa_4 \big ) & a > \kappa_a
      \end{cases}
      \right\}    
    \label{eq:pip_bcct}
\end{equation}

\section{Simulation implementation}
\label{section:computer_implementation}
To numerically solve the proposed model and evaluate the influence of the examined PIPs, we utilize the popular agent-based simulation (ABS) approach \cite{agent_based_main,agent_based_main_2}. ABS is a computational approach that is commonly used to simulate the behavior of autonomous agents within a system. In the context of botanical-epidemiological modeling, these agents represent plants within a population and their interactions with each other and with the environment. One advantage of ABS over alternative numerical approaches is that the decision-making processes and the non-symmetrical properties of individuals are easy to integrate and evaluate, making the simulation closer to realistic settings \cite{abs_intro_1,abs_intro_2,abs_intro_3}. In this section, we present the ABS implementation of the proposed epidemiological-ecological-economic model followed by the optimization of CLR control. 

\subsection{Agent-based simulation}
Each coffee tree in the population of a single coffee farm is represented by a finite state machine \cite{fsm} such that the coffee tree's state is defined to be the quantities of Eq. (\ref{eq:tree_level_system}) and the two-dimensional vector \((x, y)\) denoting its location in the physical space \cite{teddy_plants}. Formally, each coffee tree is represented by a tuple \(\tau := (H, L, I, J, U, B_p, B_r, x, y)\).

Inspired by \cite{chaos_teddy}, we implemented the simulation as follows: The entire population follows a discrete global clock, operating in rounds \(t \in [1, T]\) such that \(T < \infty\). At the beginning of the simulation, (\(t = 1\)), the  population is set to an externally defined initial condition. The ecological processes, which are set as vectors of size \(T\), are externally defined at the beginning of the simulation as well. At each round \(t\), the following three processes take place: First, the tree level epidemiological dynamics (Eq. (\ref{eq:tree_level_system})) take place for each individual tree, followed by the farm level epidemiological dynamics (Eq. (\ref{eq:farm_level_model})), thus updating the states of the coffee trees in the population. The infection dynamics are implemented in a pair-wise manner as proposed by \cite{teddy_alexi}. Then, any PIP(s) chosen for implementation are applied as discussed in the following. Finally, the accumulative profit is updated. 

\subsection{CLR control}

Recall that a coffee farm may change the implemented PIP at any point in time. As such, to allow for adjustable PIPs to be implemented, we consider the farmer's decision-making process as Markov Decision Process (MDP) \cite{markov} over $T$ rounds.
Specifically, at each time step \(t\), a given state is observed which consists of the following information: the simulation's state at round \(t\), the number of available money for PIPs implementation at round \(t\), an estimated vector of the ecological factors (\(C_T, C_H, C_R\)), donated by (\(C'_T, C'_H, C'_R\)) of size \(\psi > 0\) for the duration \([t, t + \psi]\) such that \(\forall i, j \in [t, t + \psi] \wedge a \in [T, H, R] \; : \; i \leq j \leftrightarrow ||C'_a(i) - C_a(i)|| \leq ||C'_a(j) - C_a(j)||\). 
The farmer then needs to choose an action, with the action space consisting of all combinations of the three examined PIPs (see Section \ref{sec:pips}), as defined by their combined parameter space. The farmer seeks to maximize the following multi-objective reward function:
\begin{equation}
    R(t) := w_1 \sum_{i=1}^{t}O(i) - w_2 \frac{1}{t}\sum_{i=1}^{t} R_0(i),
    \label{eq:rl_reward}
\end{equation}
where \(O(t)\) is the economic profit of the farm at round \(t\) and \(R_0 := \big ( I(t) - I(t-1) + R(t) - R(t-1)  \big ) / I(t-1)\) is the reproduction number of the CLR pandemic at round \(t\) \cite{iu_game_security,metric_paper_1,metric_paper_2,metric_paper_3,metric_paper_4}. In addition, \(w_1, w_2 \in \mathbb{R}^+\) are the weights that balance the importance of each element. 
Lastly, the transition function follows the ecological-epidemiological dynamics outlined above.

We approximate the optimal PIP policy using a deep reinforcement learning (DRL) agent. 
To train the agent, we used the ABS-RL (SiRL) approach proposed by \cite{rl_train_algo} with a wide range of initial conditions, ecological dynamics, and parameters. The agent gets as input the epidemiological state of the farm, the amount of available funds for CLR control, and the time (in days) until the end of the year, It then returns the parameters of each PIP as an action. Complete details on the DRL agent's training process are provided in the supplementary code. 

It is important to note that the epidemiological state of the farm can be obtained using one of two options: First, in the \textit{Fully observable pandemic} option, the states are obtaining accurate information on the epidemiological state of the farm. Second, in an \textit{\(\alpha\)-Partial observable pandemic} option, each state is randomly sampling from its coffee tree population at a portion \(\alpha\) and estimating the overall epidemiological state accordingly. The first and second options naturally converge for \(\alpha = 1\). Since farmers are not able to sample or monitor all of their trees at each point in time or even the majority thereof, \(0 < \alpha << 1\) seems to be a realistic setting, making the first and second options significantly different. 

\section{Analysis}
\label{sec:experiments}

In this section, use the proposed model and simulation outlined above to analyze the case of small coffee farms in Colombia. Recall that Colombia is the second largest producer of coffee worldwide and consists mostly of small coffee farms. We first outline the epidemiological, ecological, and economic parameter values used in our analysis which are based on historical records and data from prior literature. Next, we conduct three levels of analysis: baseline dynamics, parameter sensitivity, and optimal control. That is, we first establish the epidemiological and economic consequences of untreated coffee farms. Second, we determine the connection between several central parameters used by the model and the coffee farm's outputs. Finally, we estimate the ability of a farmer to sustain profit by implementing an intelligent CLR control policy. 

\subsection{Set up}
In Colombia, coffee has around an 8-month growing season depending on the ecological parameters during this time. It usually takes around 4-months to fully harvest a coffee farm, and coffee trees are harvested once per year \footnote{https://easyhomecoffee.com/does-coffee-grow-all-year-round/}. As such, we define a step in time to be \(\delta t = 1 \; day\) and \(T = 365 \; days = 1 \; year\). In particular, we assume that at \(t_{h} = 275\) the harvesting is beginning and the harvesting is equally distributed across \(T_{h} = 90\) days (i.e., 3 months), as usually the first month considered phase transfer and less actual harvesting is taken place relative to the remaining of this phase. While other countries and regions are expected to have similar coffee growth cycles, these are very sensitive to environmental differences that should be accounted for in the analysis. We plan to consider additional countries and regions for analysis in future work.  

In order to capture the CLR pandemic spread rate, we use the basic reproduction number metric \cite{brn}. In a complementary manner, to measure the economic dynamics over time, we use the normalized economic output metric, defined by \(O(t)/O(0)\). 

\subsubsection{Epidemiological settings}
In order to use the epidemiological component of the model, the mean parameter values are taken from prior literature, as summarized in Table \ref{table:tree_level_params}. According to the Colombian Coffee Growers Federation, Colombian coffee production has 5,243 coffee trees per hectare and productivity of 21.4 bags of 60 kilograms of coffee beans per hectare, assuming none of the coffee trees is infected by the CLR pandemic. We used these numbers to extrapolate the coffee berries' average growth rate\footnote{https://federaciondecafeteros.org/wp/coffee-statistics/?lang=en}. 

\subsubsection{Ecological settings}
The weather over time, \(C(t)\), can be either perfectly known to the farm owner or predicted using some forecasting model. As such, we define the \textit{Fully observable weather} and the \textit{Model-based weather} cases, respectively. In particular, we use a weather forecasting model which follows a standard SARIMA model \cite{weather_forcast} that is trained on the real daily weather (i.e., temperature, humidity, and rains) data of Colombia for the years 2011-2021\footnote{The data is obtained from \url{https://www.meteoblue.com/en/weather/archive/climatePredictions/colombia_colombia_3686120}}.  

\subsubsection{Economic settings}
The scale economy occurs in the segment where the costs behave according to the concavity property. One aspect of scale economies is the \say{0.6 rule} \cite{eco_role}. This rule refers to the relationship between the increase in production cost \( TC\) and the increase in production volume \(v\in V\) given by \((TC_v/TC_{v+1}) = (Q_v/Q_{v+1})^\alpha\), where \(\alpha\) denotes the scale coefficient. A value of \(\alpha\) less than unity implies increasing returns to scale. The value of \(\alpha = 0.6\) is often used as a rule of thumb to obtain the cost of a volume level \(Q_{v+1}\)  given the cost \(TC_v\) associated with the level of capacity \(Q_v\). The \say{0.6 rule} states that when the volume level is raised, the cost associated with the new level is less than the cost of the previous level, implying that there are cost savings from increasing capacity. This rule of thumb is a common way to estimate the cost of expanding capacity, based on the idea that when capacity is increased, the cost of the additional capacity decreases due to economies of scale. This is because, when the capacity level is increased, the fixed costs associated with higher capacity are spread out over more units of production. 

At the end of 2019, the value of the coffee crop was 7.2 trillion pesos (i.e., \$2.2 billion in the average exchange rate of the same year)\footnote{https://federaciondecafeteros.org/wp/listado-noticias/colombian-coffee-production-closed-2019-at-14-8-million-bags/?lang=en}. Hence, if we consider the increase in costs in moving from one to two hectares, assuming that the cost shown in Table \ref{table:cost_production} is for the first hectare, and the amount of coffee bags will increase from 21.4 to 42.8, when the scale coefficient \(\alpha\) is 0.6, we will get (according to the rule \((TC_v/TC_{v+1}) = (Q_v/Q_{v+1})^\alpha\)) that the total cost will increase by \$11711 to reach \$5028.8.
However, these costs are only correct for a season without a CLR outbreak. In that case, it is required to add to the cost to include CLR elimination costs. According to \cite{epidemic_outbreak}, this cost ranges between  USD 90–472/acre, with the cost of labor ranging from USD 15–20/h using a backpack sprayer to USD 40/h using a tractor sprayer.
The cost of monitoring CLR across the entire season is USD 150–175 based on 6–7 CLR surveys at a cost of USD 25/h. For the full coffee season, the total cost to manage CLR ranged between USD 450–1167 per acre (0.4 hectare). However, these costs are for Hawaii (USA). If we use the ratio between the GDP per capita between Colombia and the USA as a measure of the price differences, we will get a ratio of 8.7\%\footnote{https://data.worldbank.org/indicator/NY.GDP.PCAP.CD?locations=CO}. Hence the costs in Colombia are presumed to vary between USD 39-102 per acre, which is USD 98-255 per hectare. For simplicity, we assume that the maximum amount is for one hectare, and that the cost will decrease linearly with a factor USD 2.5 as we increase the number of hectares up to the given minimum of USD 98 per hectare.

According to Fedecafe\footnote{https://federaciondecafeteros.org/}, there are nearly 543,000 coffee-growing families in Colombia. Of these, 96\% are small (with less than 5 hectares of land), 3\% are considered medium (between 5 and 10 hectares), and 1\% are large (more than 10 hectares)\footnote{In recent years, large foreign investors have begun producing coffee in Colombia, purchasing thousands of hectares of land for production, so these figures may change in the foreseeable future.}.  

According to the U.S. Department of Agriculture (USDA)\footnote{\url{https://apps.fas.usda.gov/newgainapi/api/Report/DownloadReportByFileName?fileName=Coffee\%20Annual_Bogota_Colombia_CO2022-0010.pdf}}
, in 2021 planting density stands at 5,268 trees per hectare, with productivity of 19.3 green bean equivalent (GBE) bags per hectare, while Colombia’s current total productivity potential is estimated at 14.7 million bags with an average of around 14 million bags during the last years. Coffee bean exports represent over 90\% of total Colombian exports followed by soluble coffee and roasted coffee.
More disaggregated categories, such as the specific labour task and type of input, are also provided. When all economic costs are considered, average costs per hectare in 2015/16 were US\$3,318. Of this total, 57\% (US\$1,908) correspond to hired labour, 18\% (US\$586) to unpaid labour, 16\% (US\$519) to inputs, and 9\% (US\$305) to fixed costs. Table \ref{table:cost_production} summarizes the full economic costs. 

\begin{table}[!h]
\center 
\begin{tabular}{|p{0.7\textwidth}|p{0.1\textwidth}|p{0.1\textwidth}|}
\hline
\multicolumn{1}{|c|}{Cost Category Per Hectare}         & Value                    \\ \hline

\textbf{Paid labour}                                       &     \(1,907.92\)     \\ \hline
    Labour pruning and weeding         &     \(245.13\)                        \\ \hline
    Labour fertilizing                  &       \(75.39\)         \\ \hline
    Labour spraying                     &       \(48.99\)         \\ \hline 
    Labour harvest                    &     \(1,538.41\)      \\ \hline
\textbf{Unpaid labour}  & \( 586.11 \)     \\ \hline
    Labour pruning and weeding  & \( 79.57 \)    \\ \hline
    Labour fertilizing  & \( 27.24 \)    \\ \hline
    Labour spraying  & \( 12.11 \)    \\ \hline
    Labour harvest  & \( 467.19 \)   \\ \hline
\textbf{Inputs}                                       &     \(519.18\)     \\ \hline
    Herbicides         &     \(2.16\)                        \\ \hline
    Pesticides                  &       \(22.46\)         \\ \hline
    Fertilizer                     &       \(494.57\)         \\ \hline 
\textbf{Fixed costs}                    &     \(304.59\)      \\ \hline
    Installation costs  & \( 40.80 \)     \\ \hline
    Depreciation of machinery  & \( 112.93 \)    \\ \hline
    Opportunity cost of land  & \( 97.50 \)    \\ \hline
    Finance cost  & \( 53.36 \)    \\ \hline
\textbf{Full economic costs}  & \( 3,317.80 \)   \\ \hline
    
\end{tabular}
\caption{Average production costs per hectare in 2015/16 (in US\$). The values are taken from \cite{intro_intro_1}.}
\label{table:cost_production}
\end{table}

In Columbia, the National Coffee Growers Committee (FNC)\footnote{https://thosecoffeepeople.com/how-much-are-green-coffee-beans/} is in charge of selling the majority of commercial coffee crops, and it sets the price of coffee \say{per carga}. This term is used to refer to 125kg of parchment, pre-dry mill, presuming the coffee is in perfect condition. While the per carga green coffee price fluctuates in line with the market, at the time of writing, the domestic reference price currently stands at 2.160 Colombian pesos (approximately \$440.43 USD). However, these prices are just for standard coffee if a co-op buys directly from producers in towns, not providing any bonuses for the quality of the coffee. For high-scoring specialty coffees, their prices are based on quotes obtained from producers where a margin of between 10-30\% on top of the per carga price is added. Producers calculate these prices by factoring in operational costs, and the costs of research and development (more advanced coffee processes require many experiments to perfect).
However, for low-grade coffee beans, the price for 125 kilograms is \$60.000 Colombian pesos (approximately \$12.23 USD).

\subsection{Baseline dynamics}
\label{sec:baseline}
\begin{figure}[hbt!]
    \centering
    \includegraphics[width=0.99\textwidth]{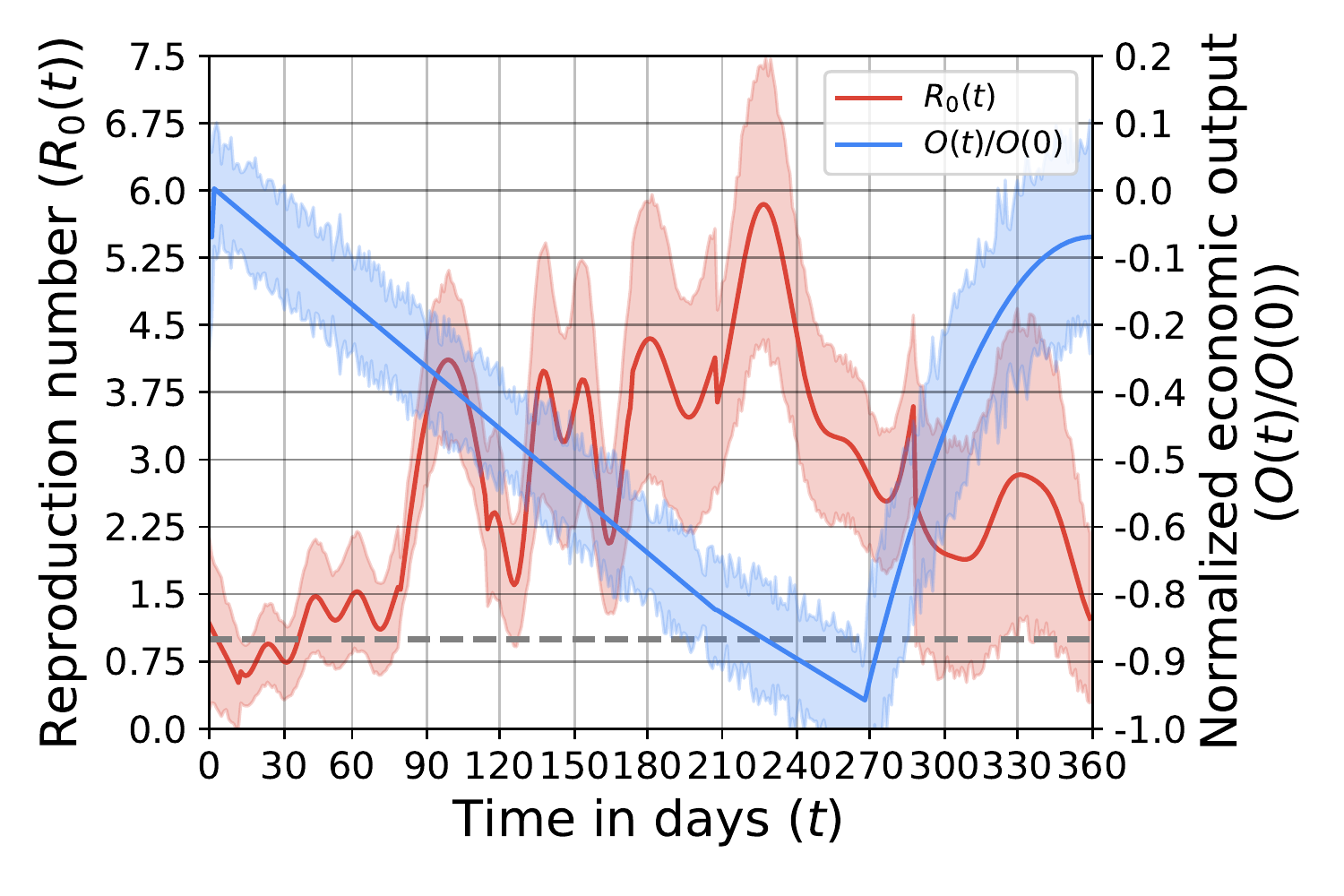}
    \caption{Baseline dynamics of the epidemiological and economic processes in a single season (one year) without any CLR control measures. The results are shown as the mean \(\pm\) standard deviation result of \(n = 10000\) repetitions divided into 100 unique farms and 100 interactions for each farm. }
    \label{fig:baseline}
\end{figure}

To start, we simulate the epidemiological and economic dynamics of the coffee farm over one season with the CLR pandemic. Using \(n_f = 100\) different farms, which differ in their size, tree density, and the initial amount of funds, we computed \(n_i = 100\) simulations each such that each one differs due to the stochastic nature of the simulation. The initial condition for the epidemiological model is set to agree with \cite{ref_4}.
Fig. \ref{fig:baseline} for each farm using the method proposed by \cite{teddy_economy}. In total, \(n = n_f \cdot n_i = 10000\) simulations are computed to capture as representative a set as possible. Fig. \ref{fig:baseline} summarizes the results where the x-axis is the time from the beginning of the simulation, starting at the end of the harvesting of the previous season. The left y-axis of the red line indicates the reproduction number (\(R_0(t)\)) and the right y-axis of the blue line indicates the normalized economic output (\(O(t)/O(0)\)). Recall that in this analysis, it is assumed that the farm owner does not implement any PIP to tackle the CLR pandemic.

One can observe that the economic output is linearly decreasing during the coffee tree growth phase as the expenses are assumed to be equally distributed. During the harvesting phase, where the coffee is sold, the economic output is increasing sub-linearly (i.e., at a rate that is less than a linear one to the source variable). Without a pandemic, the increase at this phase is linear due to the assumption the harvesting occurs at a fixed rate. However, the dead or CLR-infected trees infected during this phase reduce over time across the linear harvesting, resulting in sub-linear dynamics. One can notice that at the lowest point, the economic output reaches -0.96 which means that only 4\% of the initial funds are available at that time. Moreover, at \(t = T\), the economic output is around \(-8.5\%\) compared to \(O(o)\) which means that the farmer loses substantial money at the end of the season and cannot support the business without external help or loans. In a complementary manner, the basic reproduction number reveals an oscillating behavior due to the combined influence of the ecological signals which have such dynamics. In addition, during the dry time of the year (from roughly April to October), the basic reproduction is on average lower than that of the rainy time of the year (from roughly October to April). 

\subsection{Optimal CLR control}
\label{sec:optimal_control}
\begin{figure}[hbt!]
    \centering
    \includegraphics[width=0.70\textwidth]{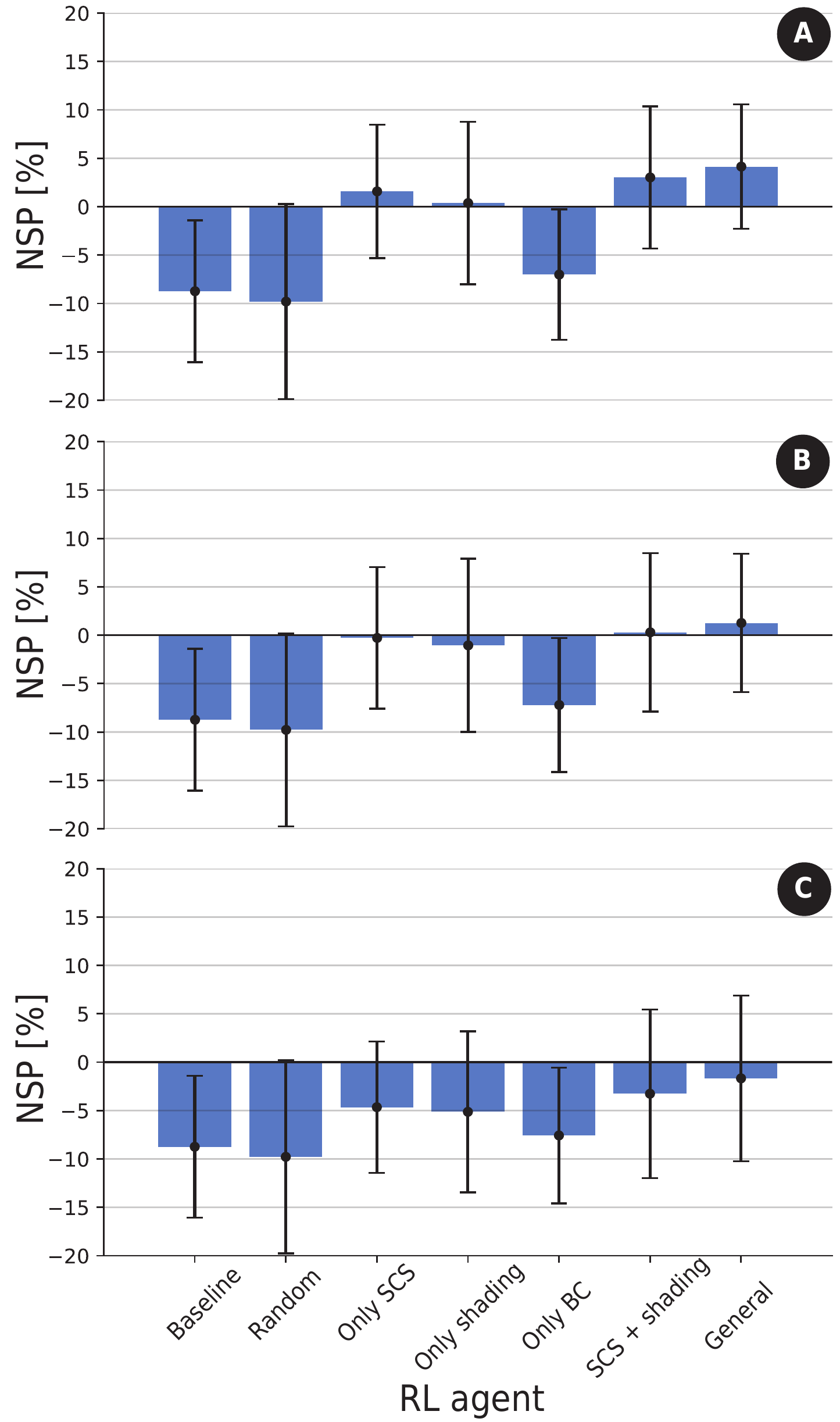}
    
    \caption{A comparison of the obtained NSP for several different PIP configurations implemented by the DRL agent. The results are shown as the mean \(\pm\) standard deviation result of \(n = 100\) repetitions. (a) Fully observable pandemic and Fully observable weather. (b) Fully observable pandemic and Model-based weather. (c) \(\alpha\)-Partial observable pandemic and Model-based weather, where \(\alpha = 0.05\).}
    \label{fig:optimal_control}
\end{figure}

To study the ability of a DRL agent to perform optimal pandemic control given limited funding, we trained the agent on \(n_{train} = 10000\) random simulations, each time with a different action space configuration as detailed next. Then, we compared the results of all agents on the same \(n_{eval} = 100\) farms, reporting the results as mean \(\pm\) standard deviation of the normalized season profit (NSP), such that \(NSP := \frac{O(t)-O(0)}{O(0)}\). We repeat this entire process for three different configurations. First, with a fully observable pandemic and fully observable weather. Second, with a fully observable pandemic and model-based weather. Finally, with a \(\alpha\)-partial observable pandemic and Model-based weather, where \(\alpha = 0.05\). For comparison, we explore seven configurations. A \say{baseline} configuration where no PIPs are utilized, a \say{Random} configuration where PIPs are utilized at random (i.e., not using the DRL agent), a \say{only SCS}, \say{only shading}, and \say{only BC} configurations where the DRL agent is able to use only the SCS, shading, and BC, respectively. The \say{SCS + shading} configuration stands for the case where the DRL agent is able to utilize both the SCS and shading PIPs. Finally, the \say{General} configuration stands for the case where the DRL agent is able to utilize all three PIPs. Fig. \ref{fig:optimal_control} shows the results of this analysis, divided into three cases. Importantly, one can notice that for the last case (c), which is presumably the most realistic one, the average NSP is negative for all action space configurations. 

\subsection{Sensitivity analysis}
\label{sec:sens}
In order to evaluate the influence of the epidemiological and economic dynamics on the farmer's NSP, we computed the model's sensitivity under the realistic assumption of \(\alpha\)-Partial observable pandemic and model-based weather agent, where \(\alpha = 0.05\). To this extent, we tested the average infection rate, average pathogen aggression, coffee cost, and labor cost as presented in Fig. \ref{sec:sens}, where the x-axis is the parameters' values and the y-axis is the NSP. The results are shown as mean \(\pm\) standard deviation of \(n = 100\) different coffee farms. We determine the ranges of the parameters' values as follows. For the epidemiological parameters, we used 10\% inspired by \cite{teddy_plants}. The economic parameters values are based on the prices of a carga (125 kg bag of green coffee) between August 2021 and August 2022 (the last month for which data was found on the website of the National Coffee Growers Committee\footnote{Coffee Statistics - Federación Nacional de Cafeteros (federaciondecafeteros.org)}), the monthly price change ranges from a 10\% decrease to a 12\% increase. Furthermore, and in recognition of the importance of labor cost to profit, we will examine changes in the cost of employing workers within a range of 10\% yearly change. The base cost of 1,907.92\$ is taken from table \ref{table:cost_production}.

\begin{figure}[hbt!]
    \begin{subfigure}{.5\textwidth}
        \includegraphics[width=0.99\textwidth]{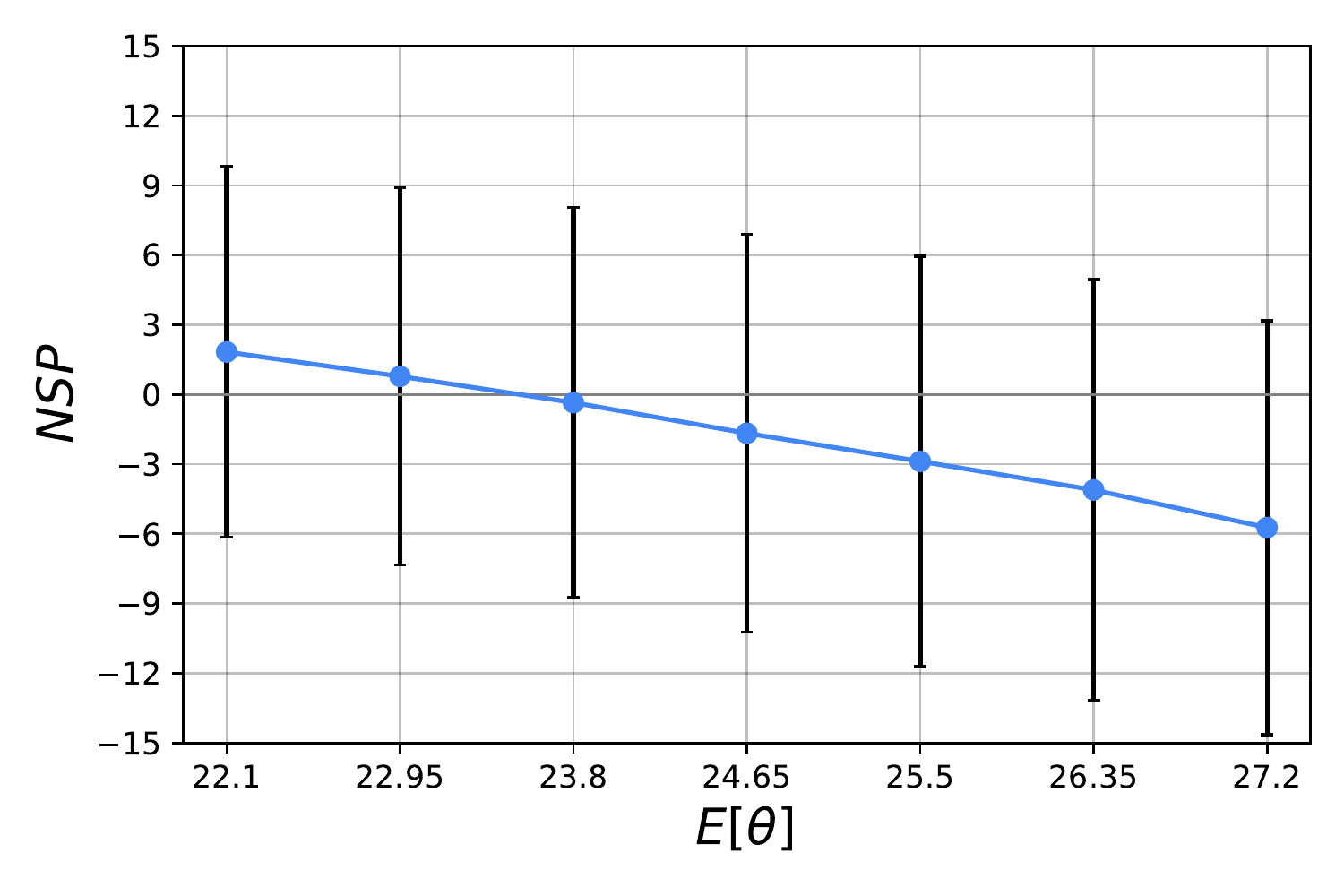}
        \caption{Average infection rate (\(E[\theta]\)).}
        \label{fig:sens_infection_rate}
    \end{subfigure}
    \begin{subfigure}{.5\textwidth}
        \includegraphics[width=0.99\textwidth]{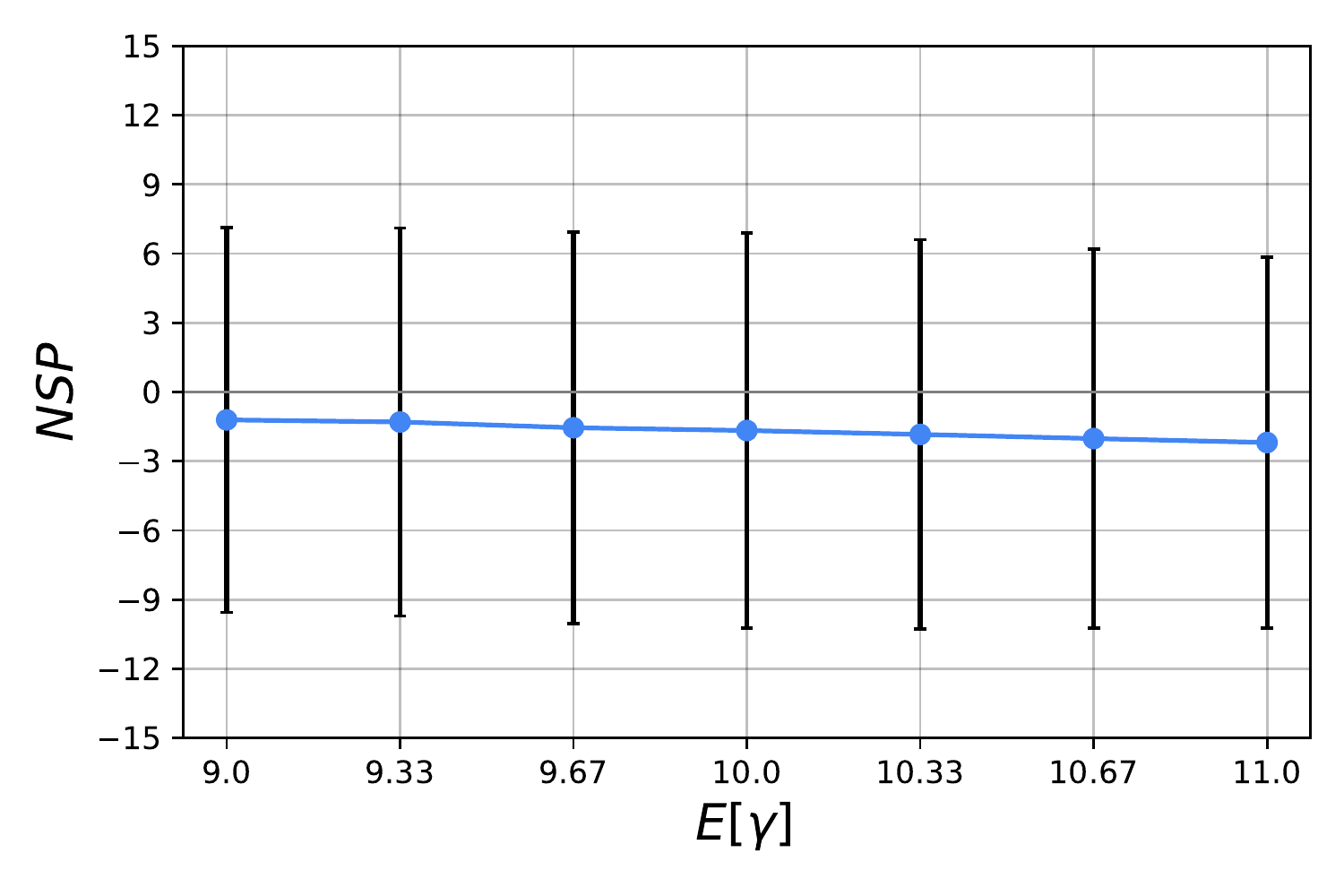}
        \caption{Average pathogen aggression (\(E[\gamma]\)).}
        \label{fig:sens_pathogen_aggretion}
    \end{subfigure}
    
    \begin{subfigure}{.5\textwidth}
        \includegraphics[width=0.99\textwidth]{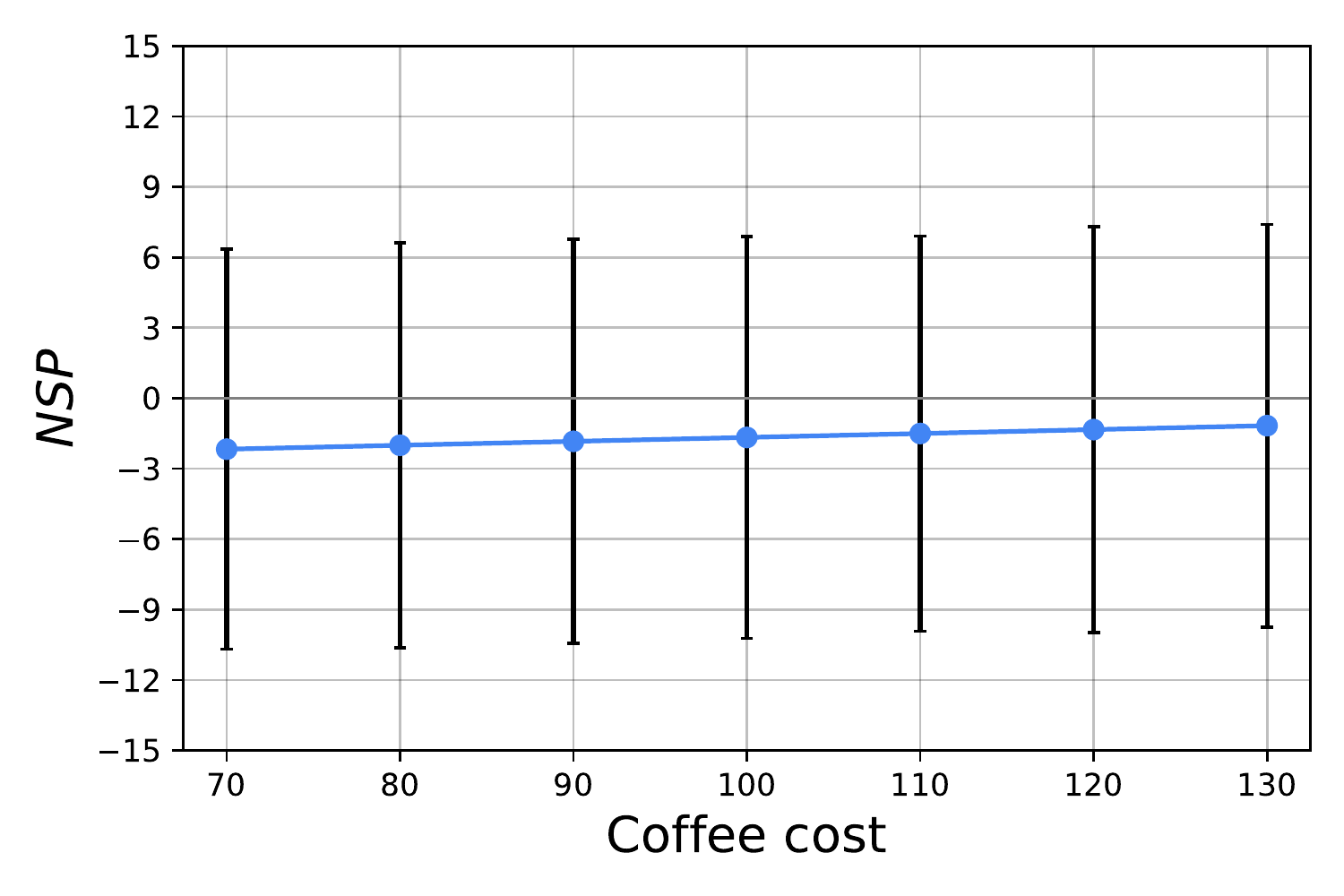}
        \caption{Coffee cost.}
        \label{fig:sens_coffee_cost}
    \end{subfigure}
    \begin{subfigure}{.5\textwidth}
        \includegraphics[width=0.99\textwidth]{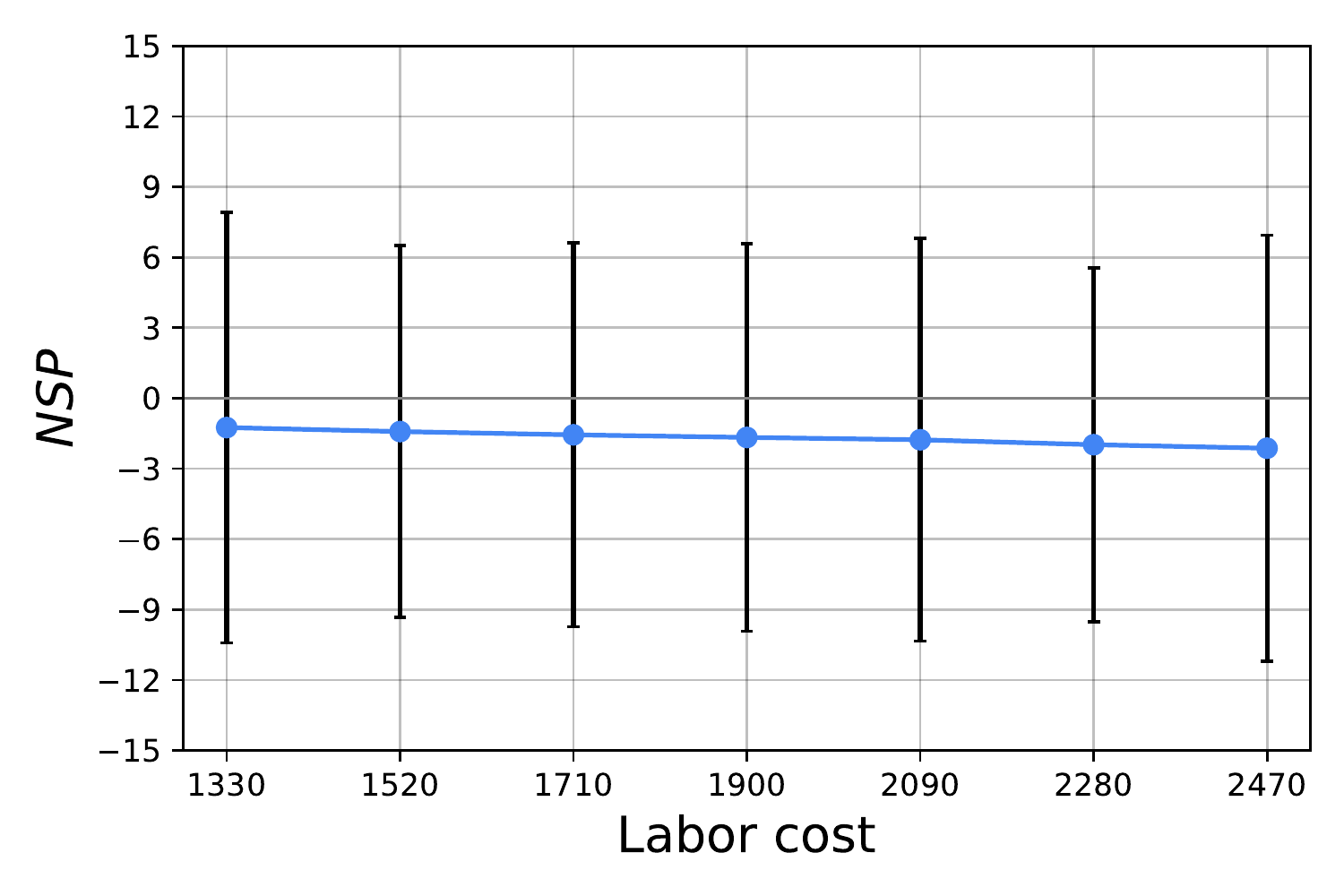}
        \caption{Labor cost.}
        \label{fig:sens_labor_cost}
    \end{subfigure}
    
    \caption{Sensitivity analysis of the NSP with respect to the main epidemiological and economic properties of the system. The results are shown as mean \(\pm\) standard deviation of \(n = 100\) repetitions for the case of \(\alpha\)-Partial (\(\alpha = 0.05\)) observable pandemic and model-based weather DRL agent.}
    \label{fig:sensitivity_all}
\end{figure}

\section{Discussion and Conclusions}
\label{sec:discussion}
In this study, we have developed a high-resolution mathematical model and a respective simulation to analyze whether an effective CLR management policy could bring about sustainable profit for small-size coffee farmers. Our analysis, which is based on real-world data, prior literature, and advanced modeling and optimization techniques, suggests that the answer is, unfortunately, mostly negative. 

Starting with a baseline analysis, our analysis established a connection between the pandemic's spread over time and its influence on the economic output. On average, Fig. \ref{fig:baseline} shows that if farmers do not implement any CLR management at all, they lose around \(8.5\%\) of their funds on average. In this scenario, 87\% of the farmer will not even break even at the end of the year. These results seem to align with historical data \cite{ref_4,ref_5}. 

When an advanced artificially intelligent agent is adopted for optimizing the economic outcomes by implementing a CLR management policy in a realistic setting, most farmers remain in losses, as shown in Fig. \ref{fig:optimal_control}, panel (C).
In the hypothetical settings where one can fully observe the epidemiological state and/or perfectly predict the temporal environmental dynamics, the agent can bring about some, albeit minor, profit when using all three PIPs, as shown in panels (A) and (B). 

Based on a sensitivity analysis, our main results seem to be robust with reasonable parameter changes. 

Taken together, our alarming results should be considered by farmers and other stakeholders as a call for action. Specifically, according to World Coffee Research, 1.7 million coffee workers lost their jobs due to CLR pandemic thus far. Our analysis seems to indicate that this is not a passing phase but, in fact, there is an inherently complex issue in the current economical-epidemiological dynamics underlying the existing coffee market. Without direct action, small-size coffee farms will likely collapse.  

Nonetheless, the proposed model is not without limitations. Unlike most cases in reality, we assume that farmers finance the cost of the PIPs from their available money and do not obtain loans or financial aid from the government. Therefore, in future work, one can introduce additional factors such as budgets, taxes, and interest to make the model even more realistic and include a more rigorous economic model that includes several types of players with different objectives and action spaces \cite{chaos_teddy}. In addition, recent developments of more rust-resistant coffee trees can change significantly the CLR pandemic spread. Taking into consideration the different coffee tree types and their relevant properties will allow one to investigate this novel course of action. Another possible extension of the proposed model is to integrate more sophisticated sampling strategies compared to the random sample of the population we used \cite{final_cite}. In addition, this study focuses on a CLR of a single season (or year). However, PIPs applied in one year can have a significant effect on years to follow. Hence, exploring the effect of extending the simulation for multi-year and multi-farm settings may shed more light on the CLR pandemic dynamics. Finally, the pathogen at the root of the CLR pandemic naturally mutates, such as other pathogens which results in a multi-strain botanical pandemic. Extending the proposed epidemiological sub-model proposed in this work to describe multi-strain or even multi-mutation dynamics such as \cite{teddy_multi_strain}, would result in a more realistic representation of the modern rust pandemic. 
Finally, as the obtained results are not linearly scalable, bigger farms do not necessarily mean that the results will be similar or even in the same direction of change. Hence, an investigation into other market settings necessitates the re-configuration and analysis of the proposed model.

\section*{Declarations}
\subsection*{Funding}
This research did not receive any specific grant from funding agencies in the public, commercial, or not-for-profit sectors.

\subsection*{Conflicts of interest/Competing interests}
The authors have no conflicts to disclose.

\subsection*{Data availability}
All the data used as part of this research is publicly available and the relevant sources are cited.

\subsection*{Author Contributions}
Teddy Lazebnik: Conceptualization, Methodology, Software, Formal analysis, Investigation, Resources, Data Curation, Writing - Original Draft, Visualization, Project administration. \\ Ariel Rosenfeld: Validation, Writing - Review \& Editing. \\ Labib Shami: Conceptualization, Methodology, Data Curation, Writing - Original Draft. 

\printbibliography

\end{document}